\documentclass[journal=dce]{CUP-JNL-DTM}%

\usepackage{graphicx}
\usepackage{multicol,multirow}
\usepackage{amsmath,amssymb,amsfonts}
\usepackage{mathrsfs}
\usepackage{amsthm}
\usepackage{rotating}
\usepackage{appendix}
\usepackage{ifpdf}
\usepackage[T1]{fontenc}
\usepackage{newtxtext}
\usepackage{newtxmath}
\usepackage{textcomp}
\usepackage{xcolor}
\usepackage{lipsum}
\usepackage[colorlinks,allcolors=blue]{hyperref}
\usepackage{booktabs}
\usepackage{cleveref}
\usepackage{csquotes}
\usepackage{subcaption}

\graphicspath{{figs/}}
\theoremstyle{definition}

\numberwithin{equation}{section}

\jname{Data/Math}
\articletype{ARTICLE TYPE}
\jyear{YEAR}

\begin{document}

\begin{Frontmatter}

\title[Article Title]{Design of policy digital twins incorporating multi-level agent based modelling }


\author[1,2]{Matt Tipuric}
\author[1]{Alejandro Beltran}
\author[1,3]{Nick Malleson}
\author[1,4]{Dani Arribas-Bel}
\author[1,2]{David Wagg}

\authormark{Author Name1 \textit{et al}.}

\address[1]{\orgname{The Alan Turing Institute}, \orgaddress{\city{London}, \postcode{NW1 2DB}, \country{United Kingdom}}}

\address[2]{\orgdiv{School of Mechanical, Aerospace and Civil Engineering}, \orgname{The University of Sheffield}, \orgaddress{\city{Sheffield}, \postcode{S10 2TN}, \state{South Yorkshire},  \country{United Kingdom}}.}
\address[3]{\orgdiv{School of Geography}, \orgname{University of Leeds}, \orgaddress{\city{Leeds}, \postcode{LS2 9JT}, \state{West Yorkshire},  \country{United Kingdom}}.}
\address[4]{\orgdiv{Department of Geography and Planning}, \orgname{University of Liverpool}, \orgaddress{\city{Liverpool}, \postcode{L69 7ZX}, \state{Merseyside},  \country{United Kingdom}}.}

\authormark{Matt Tipuric et al.}




\abstract{Digital twins are used across many industries to enable better decision making. However, while policy makers at all levels (including city, national and supranational scales) have expressed a desire to integrate digital twins into their workflows, this adoption has been slow to materialise. In this paper, we discuss the key issues associated with policy digital twins, and the ways in which they differ from, and are similar to, their counterparts in other areas. We describe how multi-level agent based modelling can be used within policy digital twins to include the effects of human behaviours on outcomes; an aspect that is often largely overlooked. We also describe how digital twins can be designed for policy use cases, and present as a case study the design of a policy digital twin incorporating multi-level agent based modelling to aid a UK city council (local authority) in delivering energy transition policy. After describing both the design method used and the resultant digital twin, we discuss the effectiveness of both, as well as how the ways in which different contexts might shape the future architecture of the digital twin. 

}


\end{Frontmatter}

\section*{Impact Statement}

Designing digital twins for policy use cases is important both for policy makers, and for affected citizens. Connecting models to the systems they represent (e.g. cities, countries etc.) with real-time data inputs and outputs supports better, more timely policy making, allows policies to adapt more easily, and better involve citizens in the policy making process. In this paper we describe how multi-level agent-based modelling can be helpful in policy digital twins, by allowing the effects of human behaviours to be better taken into account. We also present a case-study, with a city level digital twin which includes multi-level agent-based modelling to aid in policy making regarding heat pump adoption.


\localtableofcontents

\section{Introduction}


Digital twins are virtual replicas of physical systems, linked by bidirectional data flows, which to date have been developed in a wide number of domains (see, for example, \cite{Wagg20, national2023foundational}). They offer many potential benefits to policy makers for a wide range of applications and contexts. For example, they can allow for better hypothesis testing, model interactions between complex systems, track policy outcomes, and help adjust measures over time in response to changes in situation or new data. Despite this, the adoption of digital twins in policy has lagged behind other areas \citep{Richter25}, due to both technical and institutional limitations.

One key issue is that, unlike some of the major areas in which digital twinning has been developed (such as manufacturing), the systems of interest to policy makers include human actors. People interact with the physical system in complex ways, causing the cyber-physical ecosystem instead to be a human-cyber-physical ecosystem. As such, policy digital twins that do not account for humans will be severely limited in their application.

One way to include human-systems into a digital twin is to represent their behaviour using Agent-Based Models (ABMs, \citealp{epstein_growing_1996,axelrod1997complexity,crooks2018agent}). ABMs are simulation models that explicitly represent the behaviour of individual components of a system, often virtual people, and allow them to interact with each other and their environment. This makes them well suited to studying emergent outcomes in complex human-cyber-physical systems~\citep{bonabeau_agent_2002, farmer_economy_2009}. In order to capture more complex human behaviors, the ABMs can be developed as nested hierarchies. For example, a Multi-level  Agent-Based Model (MABM)~\citep{brugiere_handling_2022} could include policy-maker agents who act over a country or regional area (macro-level), within which there are multiple urban areas (meso-level), and within this there are multiple household agents who make individual decisions that affect their families (micro-level). In a MABM, interactions can occur between agents at the same scale, and between levels in the hierarchy.

In this paper, we investigate the roles of MABMs in policy digital twins through the lens of a recent case study, focused on energy transition in a UK city. The outcome of this work was the development of a digital twin demonstrator that integrates climate data and temperature forecasts with an MABM representation of the population in order to assist the local authority of the city in targeting the implementation of low-carbon infrastructure (heat pumps in this case) in both social and privately owned housing.

The remainder of the paper is structured as follows. In \Cref{sec:back} the research background regarding policy digital twins and MABMs is presented. This is followed in \Cref{sec:challenges} by a discussion of the specific challenges that policy digital twins face, including definitional, situational, and data challenges. The case study is then presented in \Cref{sec:case} including a discussion of how it links to the challenges described in \Cref{sec:discussion}, before conclusions are given in \Cref{sec:conclusion}.

\section{Background \label{sec:back}}

\subsection{Digital twins for policy \label{sec:DTfP}}
Digital twins as a concept arose out of the space engineering sector. The term was first published in the United States National Aeronautics and Space Administration 2010 Technology and Processing Roadmap \citep{NASA10}, although they were already being used in the design process for decades prior \citep{Rosen15}. The roadmap defines digital twins as: \begin{displayquote}``an integrated multi-physics, multi-scale, probabilistic simulation of a vehicle or system\dots [which] integrates sensor data from the vehicle’s on-board integrated vehicle health management system\dots forecasts the health of the vehicle/system, the remaining useful life and the probability of mission success\dots [and is] capable of mitigating damage or degradation by recommending changes in mission profile''.\end{displayquote}

\begin{figure}[htb]
\centering
\includegraphics[width=0.4\textwidth]{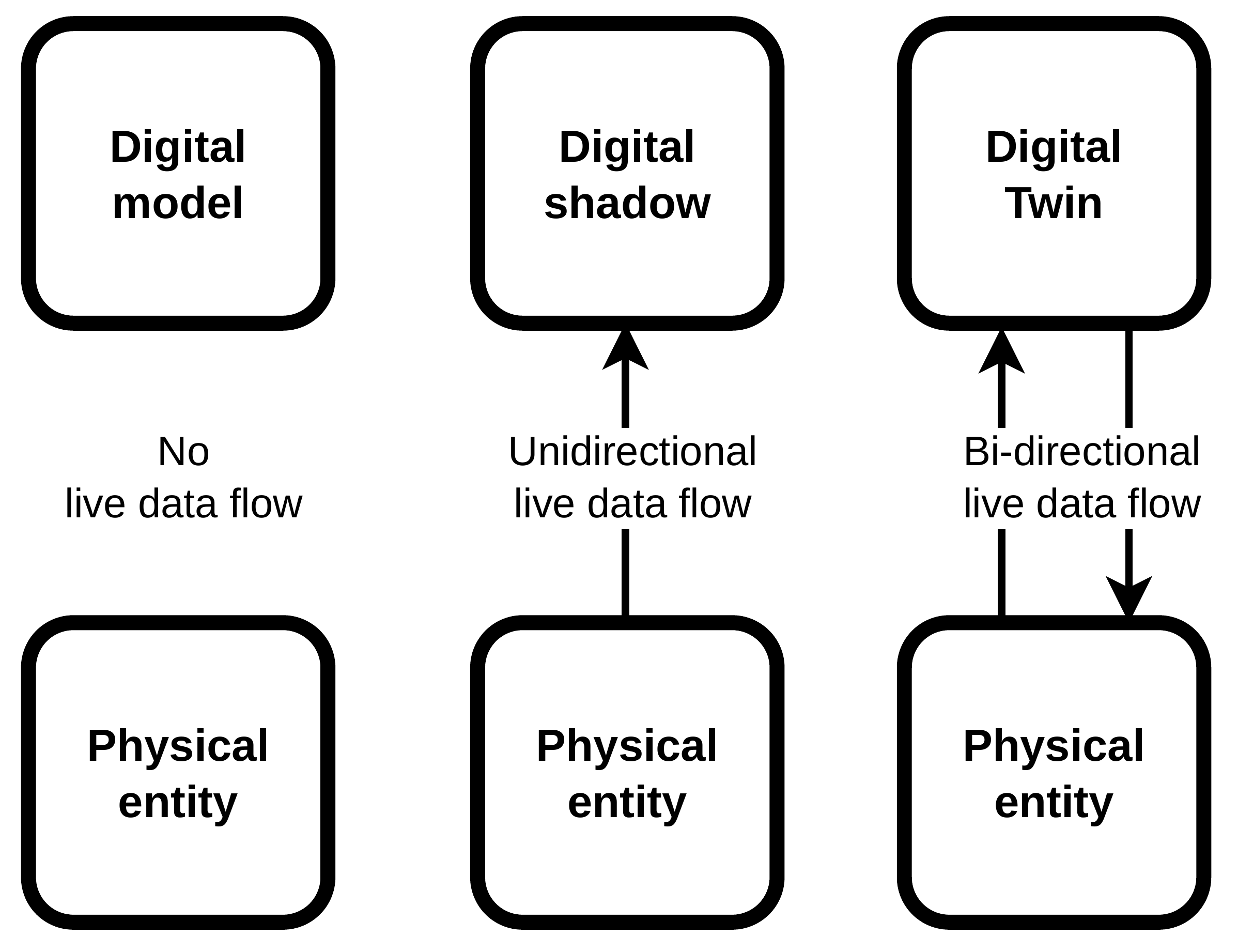}
\caption{\label{fig:block_diagram} A block diagram representation of a digital model, digital shadow, and digital twin.}
\end{figure}

In the years since then, digital twins have been adopted in many other sectors, including manufacturing \citep{Rosen15}, healthcare \citep{Ricci22}, transport \citep{DFT24}, energy \citep{MITCHELL25}, and urban planning \citep{Evangelou22}. The specific benefits digital twins offer are context specific, but include decision making, resource management, `what-if' simulations, and lifetime extension. This has required the adoption of a broader definition of a digital twin, which was provided by the International Standards Organisation (ISO) in 2023 \citep{ISO30173}:
\begin{displayquote}\textbf{digital twin (DTw)}: digital representation of a target entity with data connections that enable convergence between the physical and digital states at an appropriate rate of synchronisation. \end{displayquote}
It has become conventional to make a distinction between digital models, shadows, and twins based on the directionality of the data connections: as shown in \Cref{fig:block_diagram} models have no real-time data connection; shadows update the virtual component but not the physical component; and digital twins use bi-directional data flow to create a feedback loop. However, this classification is not without limitations: it is inherently simplistic, and the term `digital shadow' has not been adopted widely.

Digital twins have been proposed for many areas of interest to policy makers. These include the energy sector, for infrastructure planning on an international scale \citep{Teng26}, system design \citep{Granacher22}, and local level energy allocation \citep{Poshnath25}; in transport to explore urban mobility \citep{Papyshev21}, logistics \citep{Marcucci20}, and crisis resilience \citep{Tipuric25}; and in public health for interconnected healthcare \citep{Ricci22} and urban air quality \citep{Li26}. Although much of the literature focuses on city contexts (so called `smart cities' or `urban digital twins') \citep{Deng21,Herath22,Alharbi26} -- with case studies including Gothenburg in Sweden \citep{Maiullari26}, Larisa in Greece \citep{Evangelou22}, and Bertam in Malaysia \citep{Zaidi24} -- some works have also looked at the national level \citep{Rahman26,vanMeerten25} and multinational scales \citep{Hoffmann23,Wickberg25}. 

Although it is clear that there is significant interest in using digital twins for public policy, there are fewer examples of their successful adoption. \cite{Richter25} claim that digital twins remain `largely absent from administrative workflows and policymaking', despite their success in other areas. They attribute this gap to multiple social and technological causes, including technological immaturity, organisational resistance, and regulatory and financial challenges. They also identified a  sectoral imbalance in the research literature, with a predominance of use cases in the manufacturing and logistics sectors, and a dearth of telecommunications or home/commercial use cases.

\cite{Batty24} identified a key issue with the formulation of digital twins for urban planning: the lack of a `human in the loop'. He argues that social systems are a key inclusion in any concept of a city planning digital twins. By describing cities as analogous to biological processes, the links between cities and their digital twins are described as part of an evolutionary process, where both the digital and physical twins evolve to reflect one another. \citeauthor{Batty24} also describes the computational challenges associated with city digital twins, including the imperfect way in which the available data can represent the desired features, the lack of theoretical forms to represent third order interactions, and the challenges associated with visualising complex systems.   

One further complexity in discussing the landscape of policy digital twins is the heterogenous and multi-layered nature of policy-making itself. Varying, overlapping, and simultaneous responsibilities in designing policies at different levels of government make designing computational approaches difficult. Governments may compete for resources or powers to develop specific policies at multiple levels thus constraining their ability to coordinate the adoption of policy digital twins. Access to information is a particular challenge given varying degrees of digitalization or state capacity to manage information systems. A possible approach for modelling the complexities of policy-making is through agent based modelling.


\subsection{Agent Based Modelling}

Agent-based models (ABMs) offer a means of representing systems that comprise a heterogeneous collection of autonomous, interacting individuals \citep{gilbert_agentbased_2008}. Rather than describing a system through aggregate approaches, where individuals are necessarily grouped based on similar characteristics, an ABM specifies the behaviour of its constituent `agents' --- which might represent people, households, vehicles, or entire organisations --- together with the rules that govern how they act and interact with one another and with their environment~\citep{bonabeau_agent_2002}. The system-level behaviour of interest emerges from the `bottom-up'~\citep{castle_principles_2006} as a result of the many individual decisions and interactions taking place. This capacity to link micro-level behaviour and macro-level outcomes is the approach's defining strength, and makes ABMs a natural way to represent the human element in digital twins.

ABMs have been applied across a wide range of policy-relevant domains, including epidemiology and the spread of disease \citep{shook_investigating_2015}, the management of invasive species and other ecological systems \citep{anderson_geographic_2020}, managing residential segregation \citep{benenson_minority_2011}, the reliability of public-transport networks \citep{kieu_stochastic_2019}, national-level COVID policy decisions~\citep{oswald_agentbased_2024}, evacuation during natural disasters \citep{jumadi_estimating_2020}, and even modelling entire national economies \citep{poledna_economic_2023}. Their adoption in these areas reflects both the increasing availability of granular data about individual behaviour and a growing recognition that many policy-relevant phenomena are fundamentally driven by the decisions made by interacting individuals.

\textit{Multi-level} agent-based models (MABMs) are a class of ABM that move beyond a single micro-macro causal link to include interactions across multiple different levels (i.e. micro-meso-macro). They are particularly well suited to policy digital twins because, by explicitly representing interactions across multiple scales, they can, for example, link physical systems (macro), decision makers (meso), and individuals (micro) . Realising the potential of (M)ABMs is not without difficulty, however. ABMs can be computationally expensive and can contain a large number of parameters which makes them notoriously difficult to calibrate and validate~\citep{lee_complexities_2015, heppenstall_future_2020}. These challenges are compounded in a digital-twin setting, where the model must remain aligned with a continuous stream of observations from the real system --- a problem we return to in Section~\ref{sec:data_validation}.

\section{Challenges of policy digital twins \label{sec:challenges}}

Although the design of digital twins is rarely simple, the nature of policy work presents a distinct portfolio of challenges. While the challenges typically overlap with each other, we can broadly group them into three categories: definitional and conceptual challenges; challenges related to situational heterogeneity and policy ecosystems; and data, verification, and validation challenges. In the rest of this section, we go through each of these in turn and describe how they apply to policy digital twins.

\subsection{Definitional and conceptual challenges}

As highlighted in \Cref{sec:DTfP}, there are multiple competing definitions of what constitutes or does not constitute a digital twin. Although the standard \citet{ISO30173} acts as an authoritative standard, this standard is broad enough to allow multiple interpretations. In addition, different interpretations of the standard can be adopted by different governments, advisory bodies, and industry groups. While any discussion of definition might seem pedantic and academic, such criticism misses the real world implications of the confusion this causes. `Digital twin' solutions offered to policy makers range from validated models, through data dashboards, to fully automated control style systems. Many policy makers are constrained by rules designed to avoid misuse of public money; in this context, the choice of an overly constraining definition might disallow the purchase or funding of otherwise useful solutions, while an overly broad one might cause for a confused tender process. In addition, definitional confusion makes it hard for policy makers to learn lessons from each other, as one digital twin project might have entirely different qualities to another.

One compounding factor is that the nature of policy work makes it ill-suited to digital twin definitions that are suitable in other contexts. One crucial factor highlighted by \cite{Batty24} is the impact of human actors (i.e. people) on the rest of the system (or eco-system), whether that system is a city, country, collection of countries, or otherwise. There is an inherent pre-assumption of automation in digital twin definitions, that is rooted in the original NASA context and is also contained within more recent schemes \citep{Wagg20}. In other contexts, it might be suitable to abstract away the human element; for example, the highly regimented nature of factories or rail networks reduce the impacts of human decisions on the rest of the model  (or reserve them to some separate supervisory structure). However, as \cite{Batty24} notes, humans are inherently entangled in the systems we seek to make policy digital twins of --- they exist as part of the physical twins, in the data and action links, as part of the supervisory layer, and within the policy superstructure that constrains the twin. 

\begin{figure}[htb]
\centering
\includegraphics[width=0.8\textwidth]{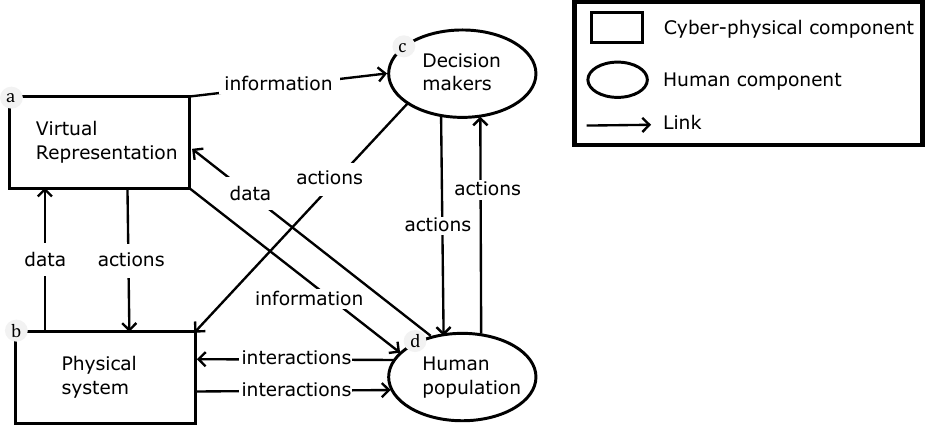}
\caption{\label{fig:HIL} Representation of a policy digital twin with humans in the loop showing the additional possible interaction between elements. Note that not all realisations of a policy digital will include all of these links.}
\end{figure}

\Cref{fig:HIL} shows how we have included humans in the loop when developing the policy digital twin. For the kinds of system of interest in this work --- whether city, country, or any other form cyber-physical-human system --- we can group the human element into two classes: ``policy makers'', such as local or national government, and ``human population'', e.g. citizens or other users of the system. In addition to the classical representation of a digital twin (cycle a$\rightarrow$b$\rightarrow$a in \Cref{fig:HIL}), we gain seven new plausible connections between elements. The human population can interact with the physical system in a reciprocal manner (d$\rightarrow$b$\rightarrow$d); for example, a population might change their travel patterns based on the design of a transport network, which in turn affects the load on the network. Decision makers can take actions on the population (c$\rightarrow$d) in the form of passing laws, while the population can affect decision makers (d$\rightarrow$c) through means such as democratic actions. Decision makers might seek information from the virtual representation of the system (a$\rightarrow$c) and use this to perform actions on the system (c$\rightarrow$b). In addition, the goal of the digital might be to provide the population with information (a$\rightarrow$d), as with weather forecasts or emergency alerts, and we might design the twin to include data from the population (d$\rightarrow$a), either directly or indirectly (such as through social networks).

\begin{figure}[h!]
    \centering
    \begin{subfigure}{0.2\textwidth}
    \includegraphics{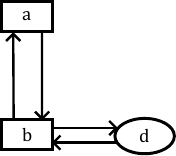}
    \caption{\label{fig:HIL1} }
    \end{subfigure}
    \hfill
    \begin{subfigure}{0.2\textwidth}
    \includegraphics{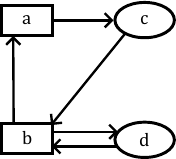}
    \caption{\label{fig:HIL2} }
    \end{subfigure}
    \hfill
    \begin{subfigure}{0.2\textwidth}
    \includegraphics{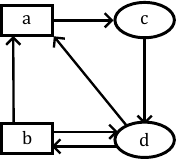}
    \caption{\label{fig:HIL4} }
    \end{subfigure}
    \hfill
    \begin{subfigure}{0.2\textwidth}
    \includegraphics{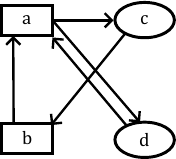}
    \caption{\label{fig:HIL3} }
    \end{subfigure}

    \caption{Four plausible policy digital twin systems, including (a) fully automated policy making, (b) government-in-the loop policy making, (c) government-in-the loop governance, and (d) participatory design digital twin.}
    \label{Fig:HIL_mult}
\end{figure}

Considering policy digital twins using the framework of \Cref{fig:HIL} helps us to reduce potential ambiguity. The additional connections afforded by including humans mean that there are now multiple ways to form a closed looped system, and a selection are shown in \Cref{{Fig:HIL_mult}}. Of these \Cref{fig:HIL1}, which represents a fully automated digital twin, is arguably closest to a `classical' digital twin of the kind shown in \Cref{fig:block_diagram}. However, in many practical cases, it is necessary and desirable to have humans in the decision making loop, as shown --- for example --- in \Cref{fig:HIL2}. If the goal is to manage the behaviour of the human population, rather than the physical infrastructure, then it is also necessary to have a data flow from the population, as shown in \Cref{fig:HIL4}. Finally, \Cref{fig:HIL3} represents the kind of digital twin used in participatory design \citep{Maiullari26}, where the twin is used to elicit feedback from the population on planned changes to the physical system.

\Cref{Fig:HIL_mult} does not represent an exhaustive set of possible digital twin layouts. In fact, the types of system of interest to policy digital twins (such as cities) can be better described as interconnected ecosystems, with significantly more complexity than the systems shown here. Nevertheless, representing digital twins as shown in \Cref{{Fig:HIL_mult}} allows us to better classify them and thus compare them more easily.

A second, related but distinct, factor is the nature of the timescales relevant to policy making, which sit between the month-scale for minor policies and guidance to decade-scale for major policy overhauls. This makes the policy unalike to any of the other contexts in which digital twinning has been applied. Even when policy makers are required to act at shorter time-scales, such as in disaster response, this work is qualitatively different to policy work, being an implementation of existing policy, rather than the creation of new policies. This increase in the length scale substantially reduces the benefits offered by holistic automation, as delivery bottlenecks are caused by other parts of the system, such as review, democratic involvement, or the creation of infrastructure. It is also not clear that such holistic automation would inherently lead to better results as, while speed of delivery can be an important factor, these lags elsewhere in the system make initial quality more valuable. Automation has a place in policy making and delivery but only in as far as it is in service of improving the policy itself.

\subsection{Situational heterogeneity and policy ecosystems}

Definitional challenges are not the only cause of difficulty in comparing one digital twin project with another. The heterogeneity of the political landscape creates a unique context for each project, meaning a detailed analysis is required to make any reasonable comparison. Different nations have fundamentally different governmental structures \citep{Caramani}, both nationally \citep{Castaneda18} and sub-nationally \citep{Moreta19}. Focusing on the local government, \citet{Wolman08} describes nine dimensions of comparison:  \textit{purpose}, \textit{importance}, \textit{structure}, \textit{decentralisation}, \textit{local autonomy}, \textit{local democracy}, \textit{service delivery}, and \textit{capacity and performance}. It is clear how divergence on any of these can radically impact the design of a policy digital twin must take. The governmental context inherently shapes the form of the project, even before taking into account the local geographic considerations, such as topography, climate, and access to infrastructure.

Comparative frameworks have largely focused on smart cities: \citet{Noori20} proposed an input-output approach wherein projects are described in terms of \textit{resources}, \textit{dynamic throughputs}, \textit{governance} and \textit{leadership}, \textit{application}, and \textit{externalities}; \citet{Alawadhi16} frames smart cities in terms of `smart governance' and focused on the three areas of \textit{governance models}, the \textit{roles and responsibility of officials}, and the \textit{decision making processes}; while \citet{Simonofski21} focused on the factors which shape citizen participation in smart city projects, characterising them as smart city \textit{characterisation} (i.e. the citizens' beliefs on what it means for a city to be smart), \textit{drivers}, \textit{degree of centralisation}, \textit{legal requirements}, and \textit{citizens characteristics}. At a meta level, \citet{Ward25} looked at the way smart cities compare themselves with one another, describing six different cities in terms of their \textit{smart city elements}, \textit{agencies}, \textit{policies}, and the presence of `\textit{elsewheres}' (i.e. the other cities with which they shared information).

The nature of research creates artificial groupings, further compounding the issue of heterogeneity. As discussed in \cref{sec:DTfP}, a focus on smart cities in the literature has led to a large number of work focused on the city scale (e.g. see \citet{Deng21}). However, the nature of specific local contexts makes it challenging to draw out general lessons that expand beyond the individual case studies from which they originate. Similarly, there is correspondingly less work focused on national or international scales. Furthermore, we lack a coherent method to analyse the relative similarity between two contexts. A hypothetical city local authority with expansive remit and powers might be better off learning from nation scale digital twin projects, for example, or from other city-scale projects with a similar geography. Without being able to quantify the distance between contexts, it is hard to make concrete statements in one way or another.

One re-framing of this issue is the consideration of policy makers as agents (or clusters of agents) within an ecosystem. Even an individual policy issue will involve co-operation and collaboration with multiple stakeholders, for data collection, infrastructure delivery, policy delivery, etc. The level of control the policy maker has over each of these depends on the level of centralisation; for only the most centralised and isolated contexts could the policy maker be considered to directly control every part of the cyber-physical-human ecosystem in which they operate. In addition, the complexity of the ecosystem increases further when multiple or interacting policy issues are considered.

\subsection{Data, verification and validation}\label{sec:data_validation}

Calibration, validation, and verification serve different purposes. Calibration tunes a model's free parameters to fit the available data. Validation asks whether the tuned model reproduces data it was never fitted to. Verification asks whether the code computes the equations as intended, rather than something that only looks right. In a policy setting, getting any of these steps wrong costs more as the twin's outputs steer interventions incorrectly; a miscalibrated or never out-of-sample tested model could send resources to the wrong places. As discussed earlier, because policy contexts are so varied a model that fits one city cannot be assumed to fit another, thus any claim that the approach transfers has to be earned by testing across several places. 

Compared to the issues discussed elsewhere in this section, those relating to both data and the verification and validation of models are possibly the most similar to other digital twin use cases, although they might present in different ways. \cite{Batty24} highlights the use of `inadequate proxies' to represent humans, which is a familiar issue to anyone studying observability issues \citep{Liu13}. While they certainly differ in quality, the issue of identifying a population's feelings based on social networks is akin to non-human problems such as identifying the internal properties of a manufacturing element based on its surface temperature or the health of a wind turbine blade based on the turbine's energy output. Equally, while they are incredibly important, issues surrounding data privacy are not unique to policy contexts, existing in any system where human data is required.

In addition, verifying and validating the models used in digital twins is fundamental to their usefulness in any context. However, the nature of policy work does create specific challenges relating to the impermanence of the physical twin and the timescales involved in delivery. In other contexts, the physical twin can be assumed to be a discrete object which can only transform within clear boundaries. A rocket engine might change the rate of fuel delivery, or have its components degrade, but over the lifespan of its digital twin, it will remain fundamentally the same rocket engine. It will not transform into a car engine, either suddenly or gradually, and so the digital twin does not need to be able to handle such transformations. The self governing nature of human-physical systems however means that over the required lifespan of a policy digital twin, the systems can be expected to radically change, due to changes in the population's  character, demography, values, politics, migration, or changes in the physical environment. If the digital twin is to self-adapt in response to these changes, then there needs to be some methodology to continuously re-verify/validate its constituent models in order to ensure they remain relevant and accurate.

One way to allow a model to adapt to changes in the physical system is to leverage the method of \textit{Data Assimilation} (DA). DA combines a model's estimate of the current system state with incoming observations to produce a more accurate picture of the `true' state of the system than either the observations or the model could provide in isolation \citep{talagrand_use_1991, kalnay_atmospheric_2003}. Data assimilation is well established in fields such as meteorology, where it underpins much of the improvement in weather forecasting over recent decades \citep{bauer_quiet_2015}, and has been used in digital twins of physical systems~\citep{kapteyn_probabilistic_2021, Wagg20}. However, incorporating data into agent-based models is considerably more challenging and no general mechanism for doing so yet exists. The difficulties are partly computational (the number of individual instances of a running ABM required to build a \textit{prior} estimate of the true system state can grow exponentially with the number of agents \citep{malleson_simulating_2020}) and partly methodological (ABMs are not readily amenable to Kalman filters and other proven DA methods \citep{wang_data_2015, ward_dynamic_2016}). A further, more fundamental, obstacle is the \textit{data association} problem, whereby the aggregate and indirect observations available from the real world must somehow be mapped onto the individual agents of the model \citep{lueck_who_2019}. In a policy setting these problems are only amplified by the diversity of relevant data sources and the continual, structural change of the human systems being modelled, making the alignment of a social digital twin with reality an ongoing task rather than a one-off calibration.

As noted by \cite{Batty24}, the human-physical systems for which we design policy digital twins operate in a different manner to purely physical systems. To capture this nuance, we re-label the `physical' twin to `human-physical twin' and include additional pathways to describe human-in-the-loop operation, and the additional feedback mechanisms available from the human-physical twin to the human-in-the-loop (i.e. the policy maker).

\section{Case study \label{sec:case}}
As a case study, we present the results from a demonstrator project which we conducted over a 16 month period to create a Scenario Planning Digital Twin (SPDT) for the Net Zero team at Newcastle County Council (NCC-NZ), who acted as our primary stakeholder and end user. 

The text and figures in this section have been adapted from an (as yet unpublished) report prepared for the European Space Agency (ESA), who funded the project.

\subsection{Local and policy context}
Newcastle City Council (NCC) is the local authority responsible for the city of Newcastle upon Tyne (which will be referred to from here as just `Newcastle', for brevity), which is located in Tyne and Wear (Northeast England) and has approximately 320,000 residents. As a metropolitan borough council, the council has responsibility for a broad range of public services, although many provisions (such as healthcare, welfare, and prisons) are delivered by the central UK government instead. The funding of these provisions comes through taxes on local domestic and business properties and grants from the central government \citep{PN07104}.\footnote{Councils are also allowed to recover direct costs through a limited number of fees charges for specific functions, such as licensing, but these are expected to be revenue neutral and thus not relevant as a funding mechanism.} Specific procedures, operations, and decision making for NCC are set out in the Newcastle Charter \citep{NCC_Charter}.

The central UK Government, via the Department for Energy Security and Net Zero (DESNZ) has energy policies that are broadly in line with the aspirations of the EU Green Deal \citep{DESNZ21}. In particular, the UK has specific targets to be net zero by 2050. As a result, policy makers at both the national and local levels are focused on developing policies that help achieve this target. 

The \citet{NCCNZ2030} defines the ambition of NCC in regard to climate policy, including achieving `net zero carbon’ by 2030. It focuses on three themes: Energy, transport, adaptation and sustainability. The Net Zero Team consists of a small number of policy officers who work with other officers in each of these areas (e.g., the housing team, the transport team, etc.) to support them in designing policy with this in mind. 

A key part of the UK Net Zero goals is to reduce the dependence on fossil fuels for residential heating. Newcastle is a pilot city for the Heat Network Zoning policy (part of the UK Energy Act 2023) and therefore NCC expects to be a leading Local Authority adopting relevant new technologies such as electrified heating systems for individual properties (e.g. heat pumps) as well as the use of district heat networks.  This presents a non-trivial challenge to develop policy that overcomes specific problems relating to retrofitting of older housing stock, changing patterns in energy demand and availability, and a requirement to educate end-users in the usage of the technology, amongst other things. Delivery will  require multiple years, and effectiveness will be affected by time-varying factors such as energy availability, a changing population, and local air temperature. These factors make this a highly suitable use case for a MABM-based digital twin.

\subsection{Design methodology}

\begin{figure}[htb]
\centering
\includegraphics[width=0.75\textwidth]{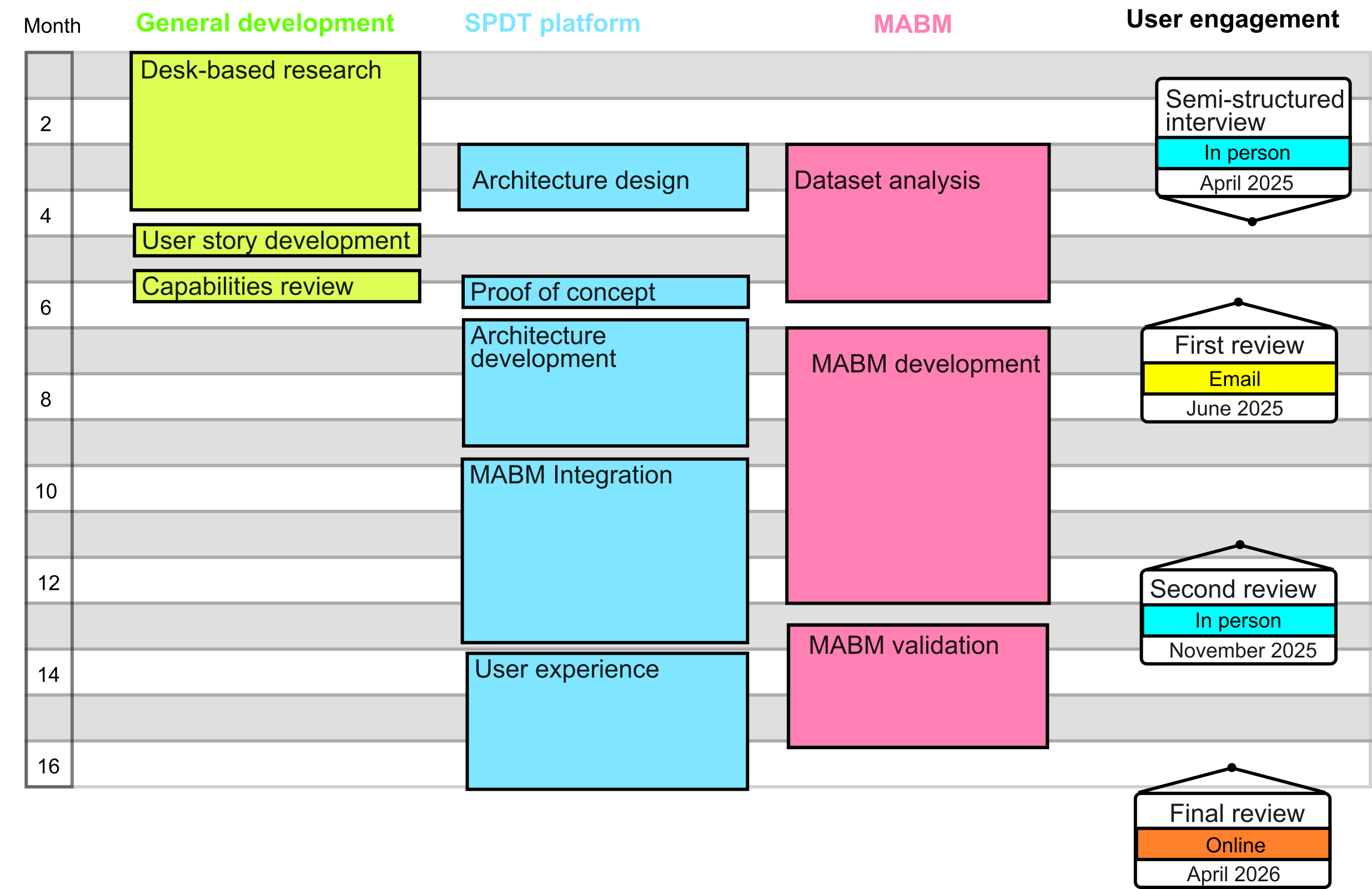}
\caption{\label{fig:tasks} Overview of the main tasks, with user engagement shown in the rightmost column.}
\end{figure}

In order to create the demonstration version of the SPDT, we adopted a design process based on agile principles (e.g., as for software design and delivery), although modified to reduce the burden on NCC as our voluntary end user. An overview of the stages of the design and delivery process is shown in \Cref{fig:tasks}. As a general approach, the MABM and SPDT were developed modularly and in parallel, with the MABM iteratively integrated into the SPDT from month 11 onward. 

User participation in the design process was achieved in the following way. The first was a semi-structured interview, conducted in person with NCC-NZ. The goal of this was to better understand the context of the user, their current requirements, and what they might want from the SPDT demonstrator. After the interview, these answers were converted into `epics’ and reviewed with the entire team. Each story was assessed for feasibility against the development timescale and team capabilities and classified as either in scope; future work; needing more research; or out of scope. After this prioritisation, a visual proof of concept was created and shared with the user via email. A second review was conducted in person after 11 months, with users given the opportunity to review progress on both the SPDT demonstrator and the ABM and provide feedback. A similar session was held online at the end of the project. 

In order to develop the MABM, it was necessary to first create a synthetic population of individuals and households from the available aggregate data. Following this, the MABM was implemented using the Mesa \citep{terHoeven25} and Mesa-Geo \citep{Wang22} Python packages. Once the basic model was implemented, the validated climate forcing was added, using the ERA5 Climate-DT datasets. The policy scenarios were then implemented by modifying the attributes of the agent at initialisation. Finally, a period of validation was conducted to assess whether the model outputs were realistic at the required scale.

The SPDT development commenced prior to the initial meeting with NCC-NZ with the design of a generic architecture for the SPDT demonstrator. After the meeting, a visual proof of concept of the platform was created to communicate the vision for the SPDT demonstrator. Once the requirements had been analyzed and prioritized, the development of the features commenced, beginning with elements that were not reliant on the MABM development. As each updated version of the MABM was developed, it was integrated into the platform.  

\subsection{MABM development}
\subsubsection{Synthetic Population}
\begin{figure}[htb]
\centering
\includegraphics[width=0.8\textwidth]{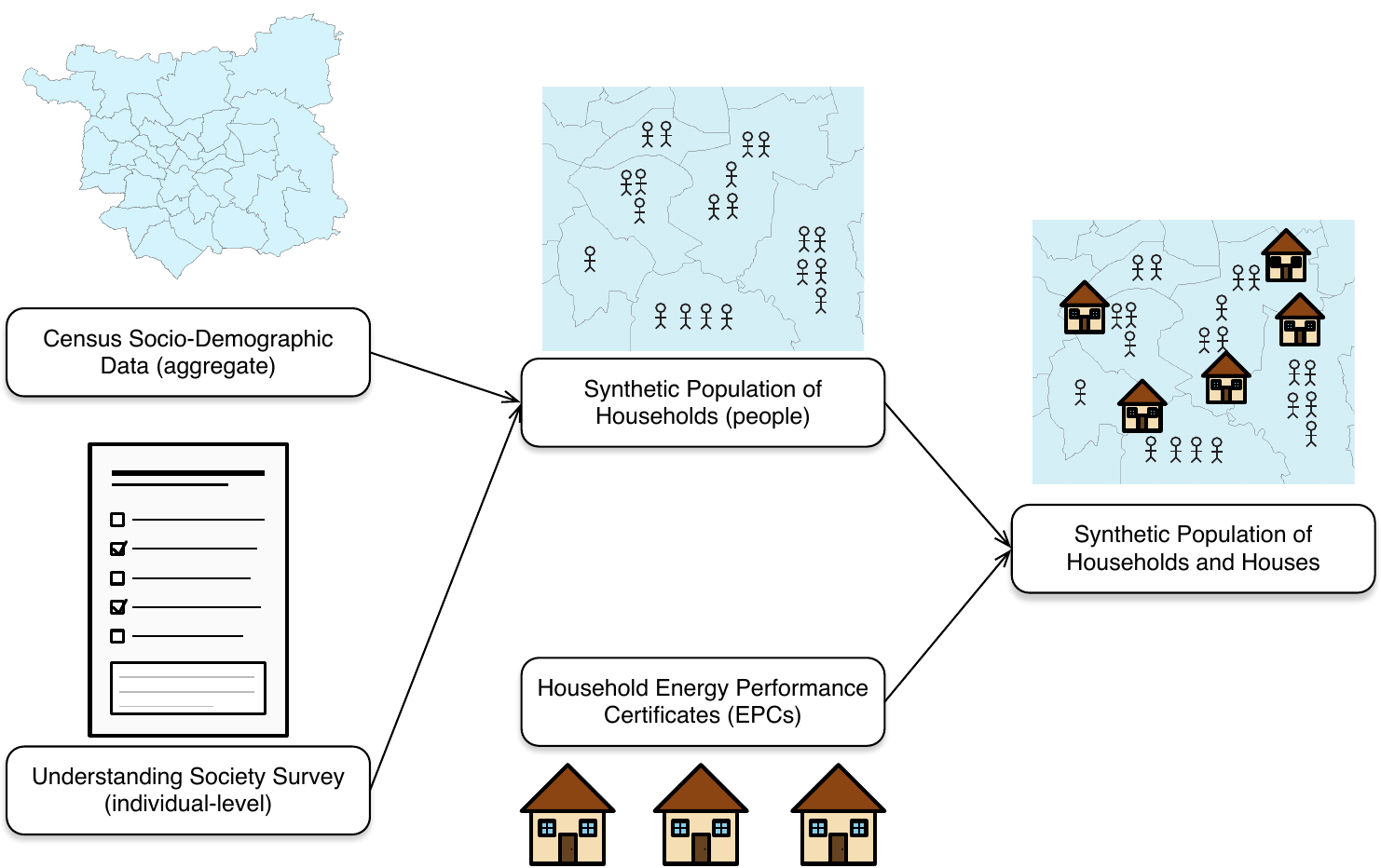}
\caption{\label{fig:synth} Data sources used to construct the synthetic population. }
\end{figure}

The MABM simulates residential energy demand at the level of individual dwellings rather than aggregate areas, so it needs both a building and the people in it for every home in the city (Figure \ref{fig:synth}). 
The synthetic people are generated using the COMPASS software \citep{hoehn_2026_18481468} by sampling from the Understanding Society survey, whose microdata are recoded to Census categories, and assigning individuals first to neighbourhoods and then to households, such that the aggregate demographic distributions reproduce the 2021 UK Census marginal distributions at LSOA level.
The buildings are instantiated from the Energy Performance Certificate (EPC) register, a statutory record of each home's thermal efficiency, and the physical attributes that determine it. From each EPC certificate we extract: property type, floor area, SAP rating, heating fuel and system, construction age. Multiple lodgements for the same property are deduplicated by Unique Property Reference Number (UPRN), keeping the most recent.  Households are then matched to individual dwellings at UPRN level by a cascading algorithm that relaxes attribute constraints only when no exact match exists; across Newcastle this delivers a 99.5\% dwelling-level match rate. All of the data used in the construction of the synthetic population are publicly available.

\subsubsection{Model Implementation}
The model is implemented in Mesa and Mesa-Geo around two agent types. \texttt{ HouseholdAgents} are the dwellings, each georeferenced within its LSOA and carrying the EPC attributes. \texttt{PersonAgents} are the occupants, each carrying income, education, a daily schedule, and household composition. A HouseholdAgent can host multiple PersonAgents, and their combined presence drives the occupancy component of demand. Each schedule type is drawn from eight archetypes derived from employment status and household composition, covering dual-earner, single-earner, part-time, retired, home-all-day, single-parent, student, and mixed households, and it governs when occupants leave and return home, how frequently they work from home, and how much load they contribute at each hour. 

At each hourly step, demand accumulates from four sources: the dwelling's structural baseline, occupancy spikes contributed by PersonAgents, climate-driven space heating, and domestic hot water. 
Climate forcing enters the model through ERA5 and Climate-DT temperature data accessed via the Destination Earth platform with each dwelling mapped to its nearest grid point. Because outdoor temperature drives the space heating and cooling terms at every step, a cold snap raises demand across all dwellings at once, but by an amount that depends on each home's fabric and occupancy. A poorly insulated, occupied dwelling in the evening peak responds to the same temperature drop very differently from a well-insulated, empty one: the multi-level architecture exists to represent the interaction between climate, fabric, and behaviour. 

Policy scenarios are set by changing agent attributes at initialisation. Heat-pump adoption switches a dwelling's fuel carrier from gas to electricity and recalculates its baseline and heating response. Retrofit measures update the envelope quality score and reduce the heating slope. A cohort selector controls which households a policy touches, leaving the rest at their baseline. The model then runs forward, and the scenario output is compared against the same baseline run to estimate demand reduction and post-hoc financial savings at user-specified unit rates. 

\subsubsection{Calibration}
Calibration sets the model's free parameters from the Smart Energy Research Lab (SERL) panel, a nationally representative smart-meter sample of approximately 13,000 UK dwellings, segmented by building type, heating fuel, and occupancy. The parameters are derived analytically from the panel and stored in a versioned configuration. The DESNZ benchmark used in the next section is deliberately held back, so that no part of it informs the fit. 

The calibration follows a strict separation between level and shape and proceeds in three stages, each targeting a different temporal dimension of demand. 

\textit{Baseline anchor}: The summer months (June–August), when space heating is effectively zero, isolate the weather-independent structural load; building-type multipliers scale it from detached houses down to flats.

\textit{Heating slope}: A regression through the origin is fitted on heating degree-days against monthly SERL consumption, restricted to the core heating months. The shoulder months are excluded because UK households turn off central heating before outdoor temperatures physically warrant it.  

\textit{Intraday shape}: SERL half-hourly data are aggregated by occupancy band (one through six-or-more) and normalised to carry shape only, redistributing energy within each day without altering the monthly total; occupancy sets both the timing and the size of the daily peak. 

Fitting level and shape seperately keeps the model accurate on the hourly, monthly, and annual scales at once. The parameter values and full protocol are available from the companion model paper [CITE: should probably arxiv it before submission]

\begin{figure}[htb]
\centering
\includegraphics[width=0.8\textwidth]{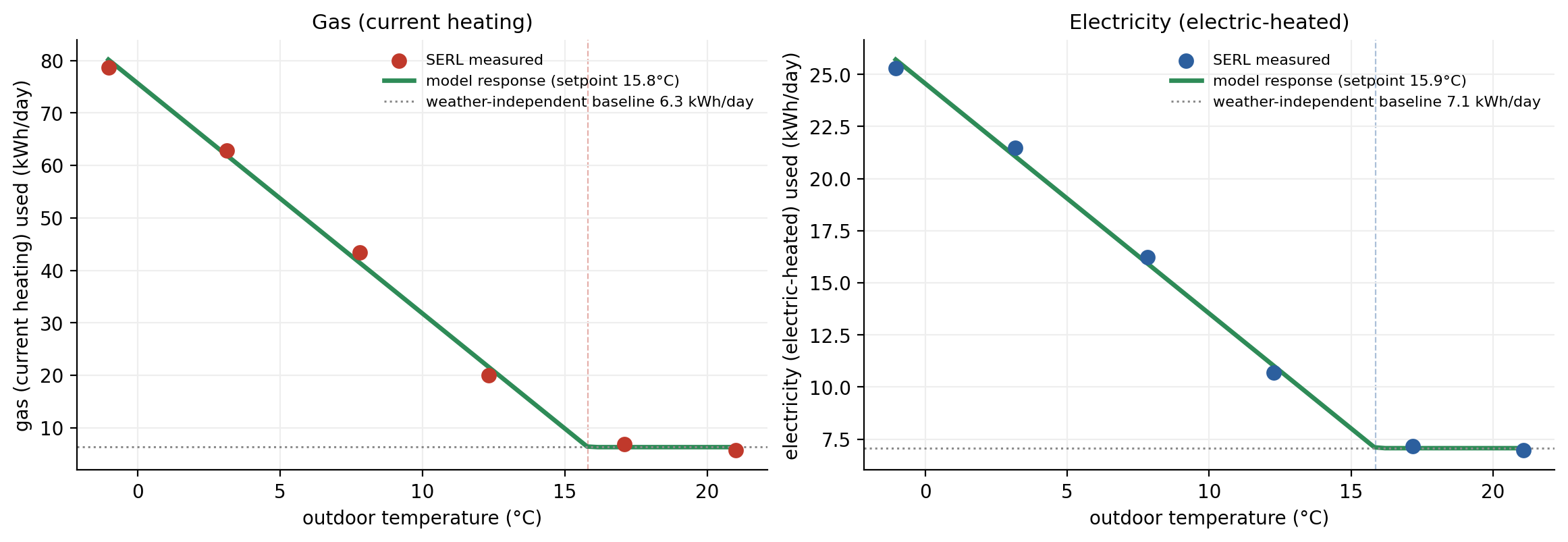}
\caption{\label{fig:dce_temp_response} Model heating response against outdoor temperature, gas-heated (left) and electric-heated (right) dwellings. Points are SERL temperature-band means; the green line is the calibrated response, not a fit to these points.}
\end{figure}

Plotting daily energy use against outdoor temperature shows what the calibration captures (Figure \ref{fig:dce_temp_response}). For both fuels, below a setpoint of roughly 16 C  each degree of cold adds a fixed increment of demand, and above it demand settles onto a weather-independent baseline. The fitted line is the model's own heating response, set by the SERL setpoint and slope, drawn over the SERL points it was fitted to, so the figure shows both what the model does and how closely it holds to the data.

\begin{figure}[htb]
\centering
\includegraphics[width=0.8\textwidth]{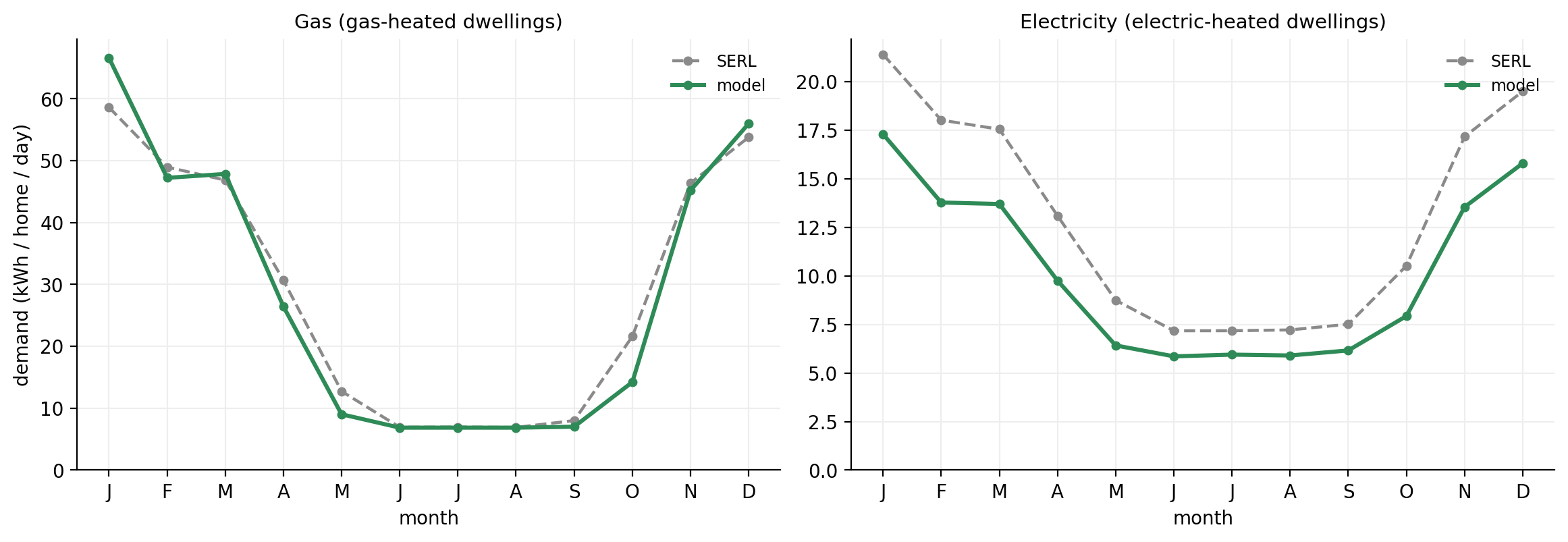}
\caption{\label{fig:dce_monthly_shape}  Per-dwelling demand by month, model against the SERL monthly mean, gas-heated (left) and electric-heated (right) dwellings, Newcastle 2023.}
\end{figure}

Over the year, the model reproduces the seasonal cycle month by month (Figure \ref{fig:dce_monthly_shape}). Gas demand in gas-heated homes tracks the SERL monthly mean from the winter peak to the summer floor. Electric-heated homes follow the same seasonal shape, though the model sits below their level throughout, it's a conservative under-representation of electric heating. 

\begin{figure}[htb]
\centering
\includegraphics[width=0.8\textwidth]{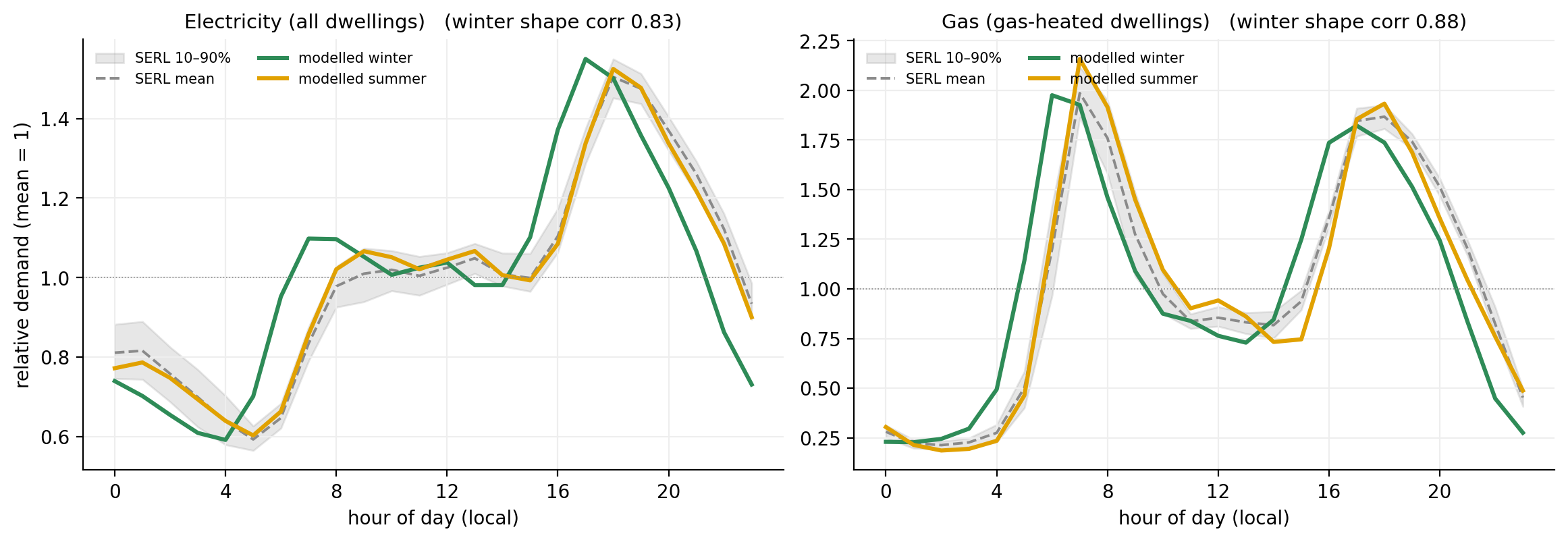}
\caption{\label{fig:dce_intraday_shape}  Modelled within-day profiles against the SERL 10–90 \% envelope, electricity across all dwellings (left) and gas across gas-heated dwellings (right), each normalised to a daily mean of 1. Winter-shape correlation with the SERL mean: 0.83 electricity, 0.88 gas.}
\end{figure}

The fit also reproduces the within-day shape (Figure \ref{fig:dce_intraday_shape}): modelled winter and summer profiles sit inside the SERL half-hourly envelope for both fuels, recovering the morning and evening peaks and the overnight trough. Matching the daily, seasonal, and annual scales at once matters, because a model can hit the annual total through compensating errors, an overstated morning peak cancelling an understated evening trough, and still be wrong about everything that shapes a load profile.

\subsubsection{Validation}

We test the model against an independent source we do not use to fit it: the Department for Energy Security and Net Zero's (DESNZ) subnational domestic consumption statistics, which report annual totals per neighbourhood. The comparison is at the annual, per-dwelling scale.

\begin{figure}[htb]
\centering
\includegraphics[width=0.8\textwidth]{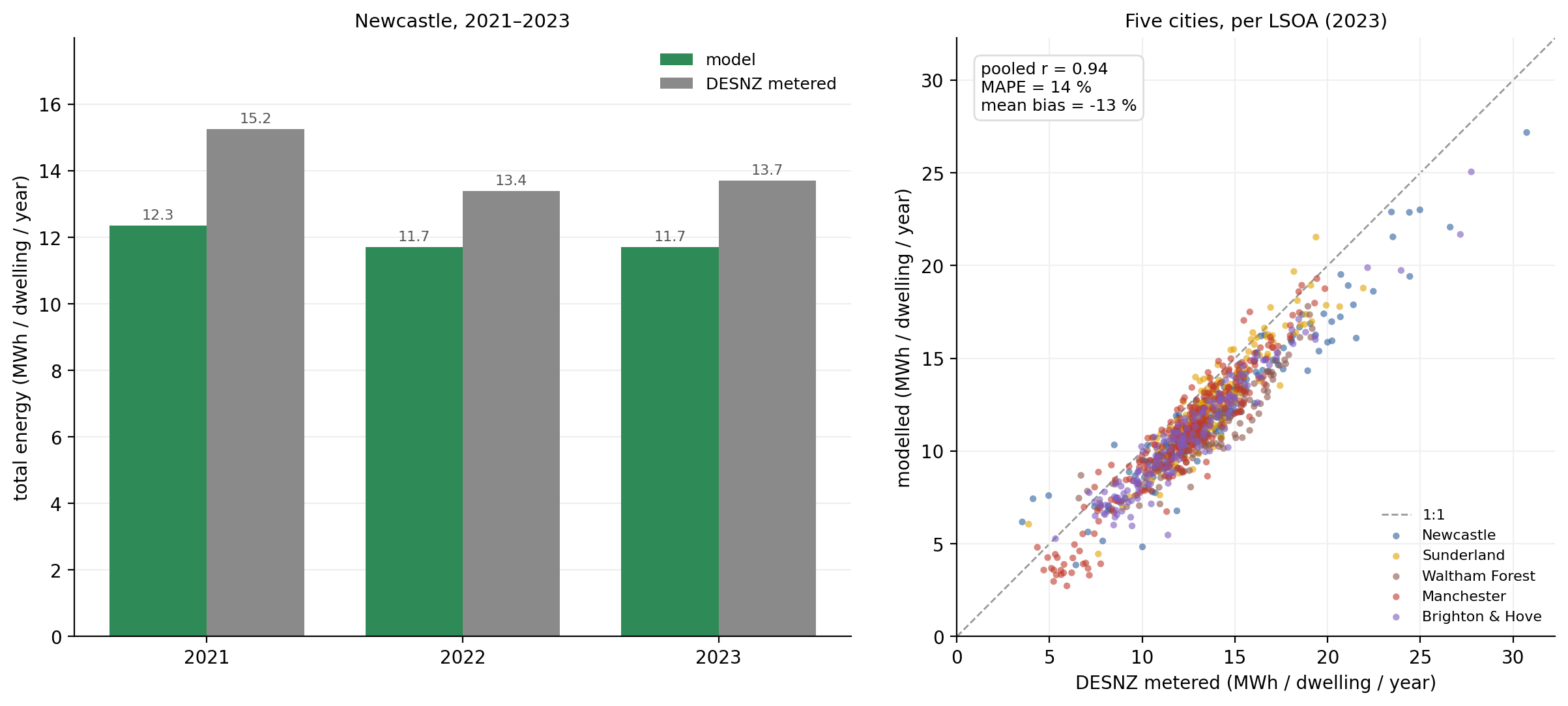}
\caption{\label{fig:dce_desnz_validation} Per-dwelling total energy, model against the DESNZ metered benchmark. Left: Newcastle, 2021 to 2023. Right: 967 LSOAs across five cities, 2023, with the 1:1 line (pooled r = 0.94, mean bias -13 \%).}
\end{figure}

Across time, we validate the model using the DESNZ 2021 through 2023 data at the LSOA level.  The model follows the direction of year-to-year change: the warmest year gives the lowest demand in both the model and the benchmark (Figure \ref{fig:dce_desnz_validation}, left). It sits below DESNZ each year by a consistent margin, wider in 2021, when pandemic-era home working lifted demand the model does not represent.

Across space, one city cannot show that the approach generalises across different contexts. We apply the same model to four more cities in England: Sunderland, Waltham Forest, Manchester, and Brighton \& Hove. Across neighbourhoods the model correlates with DESNZ at 0.90 to 0.97 (Table \ref{tab:desnz-accuracy}, showing that it reproduces the spatial pattern of demand, thus enabling targeted interventions or policy experiments. The error that remains is dominated by a single city-wide offset, as gas-driven under-count of 10 to 18\% in the same direction across every city, after correcting for that disagreement the neighborhood-level agreement is within 5 to 7\%. Figure \ref{fig:dce_desnz_validation}, right, plots model against benchmark for every neighborhood in the five cities: the points fall along the 1:1 line, with the same under-count reflected from city to city. Holding across five independent sites, the approach transfers within this class of UK local authority. 

\begin{table}[t]
\centering
\small
\caption{  MABM per-dwelling total-energy accuracy against the DESNZ benchmark, five UK cities, 2023. Correlation is per-LSOA; bias is the city-wide mean offset; scatter is the per-LSOA MAPE once that offset is removed. Gas-reliable LSOAs only (unknown gas-connection share < 15 \%).}
\label{tab:desnz-accuracy}
\begin{tabular}{l r r r}
\toprule
City & \shortstack[r]{Corr.\ with\\DESNZ} & \shortstack[r]{Total\\bias} & Scatter \\
\midrule
Newcastle        & 0.96 & $-12.3\%$ & 5.8\% \\
Sunderland       & 0.90 & $-10.3\%$ & 7.0\% \\
Waltham Forest   & 0.93 & $-18.0\%$ & 4.8\% \\
Manchester       & 0.93 & $-10.9\%$ & 6.1\% \\
Brighton \& Hove & 0.97 & $-13.1\%$ & 5.1\% \\
\bottomrule
\end{tabular}
\end{table}

\subsubsection{Verification}

Verification asks a different question from calibration and validation: whether the code computes the equations we intend. Three checks run inside the pipeline. Every parameter reported here is asserted equal to the value the fitting script emits, so the numbers in this section are the ones the engine runs. The demand decomposition closes as an identity, with the energy routed to electricity, gas, and other fuels equalling the sum of the structural baseline, occupancy spikes, space heating, and cooling, at every hour and again over the full year. And perturbing a single calibrated parameter by 10 \% moves the simulated total by the amount the closed form predicts, to within 0.2 percentage points, so the model responds to its own parameters as the equations specify.

\subsection{SPDT Development}
\subsubsection{Platform Architecture}

Architecture design was commenced before the initial meeting with stakeholders, as some of the platform requirements were known from the project brief. It was known that the final platform would need to support a user-friendly front end; be able to interface with large datasets; support some additional data manipulation tools; include some GIS interface; and support parallel development alongside the MABM. 

The platform was based on the Digital Twin Online Platform (DTOP)  \cite{Bonney22}. This is a framework to develop digital twins based on the web, using the Python Flask module for the back-end, with Bootstrap-5 for a user-friendly and responsive front-end interface. Using Python as the main language simplified the integration of the MABM – which was also built using Python -- into the SPDT architecture. Python is a widely known language and is supported by a wide range of open-source modules. Although in some cases the performance can be limited compared to other languages, it was judged to be sufficient for the scope of this project. It was anticipated that the major bottlenecks in terms of speed would occur in the running of the MABM, and in interfacing with the large datasets in the database; both cases where the choice of Python would be minimally impactful. In addition, the framework supports the use of Javascript, which can be used when front-end elements need to be more responsive. 

One of these cases where front-end performance was a concern was the GIS interface. It was known that the platform would need to be able to display the outcomes of the simulations in the form of an interactive map. This would be a new addition to the DTOP. Maplibre GL JS was chosen for this purpose, in part due to its successful use by the team in the Demoland project\footnote{\url{https://www.turing.ac.uk/research/research-projects/demoland}}. This is a Typescript library that allows geographical data in a browser to be displayed using vector tiles. It also supports the use of a database and tile cache, improving performance. 

\begin{figure}[htb]
\centering
\includegraphics[width=0.8\textwidth]{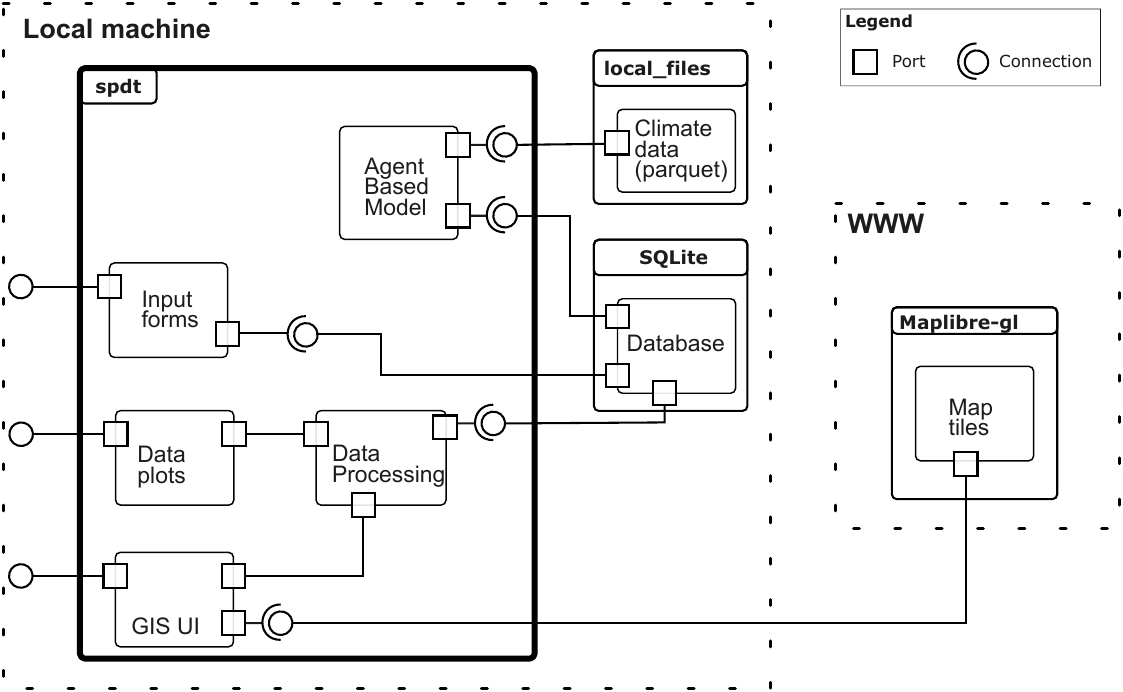}
\caption{\label{fig:architecture} Final iteration of the SPDT architecture diagram.}
\end{figure}

Following the development of the user story and the specific use case \citep{cockburn2001writing}, the full architecture construction began. It was agreed that the SPDT would not need to function as a stand-alone app but instead as a lightweight demonstrator that wired in the different component modules. This would be created with the goal of integrating those independently within the Destination Earth ecosystem as part of a planned project extension, reducing the need to reduplicate existing functionality. This meant that less focus needed to be placed on the specific implementation of the databasing and integration of the ClimateDT data, leading to the platform architecture shown in \Cref{fig:architecture}. A single SQLite general purpose database was chosen, which simplified the development of the platform. 

\subsubsection{MABM Integration}
Once a first version of both the system architecture and the MABM had been developed, the development focused on integrating the latter into the former. The initial focus was to restructure the data used and created by the MABM (i.e. input and output), as well as other user created data, to be suitable for the sqlite database (rather than the file system structure used in the MABM development). After some iteration, this led to the creation of ten database tables across three superclasses: 
\begin{itemize}
    \item Input data – static data used as an input to the ABM, defined once at initialisation, with the users granted read only access. These tables were \texttt{EPCABMdata} and \texttt{HIDPdata}, which correspond to the epc and hidp geojson and csv datasets 
    \item MABM output data – tables for each of the output types of the MABM model: \texttt{AgentTimeSeries}, \texttt{EnergyTimeSeries}, \texttt{ModelTimeSeries}. 
    \item User created data – tables for the other information defined by the user to run the scenarios: 
    \texttt{Scenario}, \texttt{ClimateModel} (not used), \texttt{Population}, \texttt{PolicyChoices}, and \texttt{Rules}. 
\end{itemize}

The MABM also required land surface data as an input. However, as SQLite is not well suited to handling large parquet files of this type, this was kept as a local file instead.
Once the initial handling of the MABM was achieved, the next step was to allow user defined scenarios to be run. In practical terms, this means allowing users to create a set of rules (\texttt{PolicyChoices}) to automatically assign a subset of agents in the simulation to the use of heat pumps. It was decided that, to streamline the interface for this demonstrator, the available flags for users to select from should be a small set of the full options. This logic required some iteration, as multiple approaches seemed plausible, balancing logic complexity with simplicity of use. Eventually, a cascading approach was selected, with users able to build a logic engine based on simple rules. To maintain consistency with the stand-alone MABM code, the adoption rate was also included, although applied globally to the whole selection, rather than as a modifiable parameter for each rule. 

\subsubsection{User experience}
Once the MABM was successfully integrated, the final step was to improve the user experience, focusing on the expected paths through the platform. The most critical of these was the ‘Create scenario’ path, as it involved a chain of relatively complex steps. To reduce cognitive burden, it was decided to split each of these steps into four sequential webpages (One each for initial setup, agent selection, and policy selection, followed by a final check and confirmation page). This approach relied on the Flask ‘session’ to store user input between each page, which means that the data would only be sent to the database once, after the user had the opportunity to check and confirm that it was correct. 

A second user experience improvement was the addition of some database management tools, which would allow the user to view and delete policy choices and scenarios in the browser, as well as clear the MABM output data for a scenario (allowing it to be rerun). Prior to these additions, these actions required either a separate database manager or the use of command line instructions, neither of which was appropriate for the target users. This was reinforced by modifying the table on the front page to indicate whether a scenario had been run or not. A lightweight data viewer for the input data was also included. Finally, a helper script was created to automate the database population with the input datasets during the installation. 

\subsection{Prototype demonstrators}
The main outcomes of this project were two prototype demonstrators: one each for the SPDT and MABM. Although the SPDT includes a version of the MABM, due to the nature of the development cycle, this is not the most recent available version. Both demonstrators are available on Github \citep{mtipuric_2026_21340939, beltran_2026_21356630}.

\subsubsection{MABM}
The MABM was instantiated for Newcastle, populating \textasciitilde97,300 HouseholdAgents from the city's EPC stock and matching synthetic occupants to each at UPRN level. The model was run hourly throughout the 2021–2023 window with ERA5 temperature forcing, calibrated against SERL aggregate statistics, and validated against DESNZ subnational consumption statistics. What follows describes how those outputs are surfaced to policy users through the SPDT and the levers that can be pulled inside it. 

The MABM is operationalised through the SPDT as a scenario engine. Users interact with the model through three coordinated controls: a cohort selector that defines which dwellings a policy applies to, a policy action that specifies what happens to those dwellings at initialisation, and a configuration override that sets uptake rates and run length. The SPDT runs a baseline and a policy version of the model over the same window and surfaces the difference in kWh totals, per-dwelling averages, and bill impacts at user-specified unit rates. 

Cohort selection is the most expressive lever. Dwellings can be masked on physical attributes (property type, floor area, SAP rating, dwelling typology), on tenure (owner-occupied, private rent, social rent), on household composition (income quintile, education level, presence of children, schedule type), or on a `top-X\%' slider that ranks dwellings according to their baseline annual consumption. These masks can be combined, allowing a policy to target, for example, low-income households in social-rent terraces, or the top decile of consumers within a specific ward. The synthetic population is what makes the socio-demographic dimensions of this targeting possible. 

Currently, three classes of policy action are supported. The first is heat-pump deployment, which flags dwellings as adoption candidates and assigns them to one of four feasibility tiers (priority, possible, difficult, non-possible). At run time, the model switches the carrier of treated dwellings from gas to electricity and recalculates the structural baseline. The second is retrofit, which sets envelope flags (cavity wall insulation, solid wall insulation, loft insulation, floor insulation, glazing) on the masked cohort. The third is occupant-side adjustment, which modifies the heating setpoint for the cohort to represent setback programs or smart-meter interventions. 

Two configuration sliders sit alongside the cohort and action controls. The adoption rate sets the share of the eligible cohort that actually receives the intervention, either globally or per feasibility tier, and the run window sets the temporal scope of the comparison, ranging from a single day for a quick smoke test through to multi-year runs out to 2039 for long-horizon planning. 

Outputs are returned to the user at three spatial scales: per-dwelling, aggregated to LSOA, and aggregated to local authority. For each scale, the SPDT reports baseline kWh, policy kWh, and the difference between them, separated by fuel carrier. Per-dwelling outputs feed the geographic visualisations that allow users to see where savings concentrate; LSOA aggregates are the natural comparison point with the DESNZ statistics local authorities currently plan against; and local authority totals are the headline figures used to compare scenarios against citywide policy targets. Bill impacts are derived after each run by applying configurable unit rates to the modelled energy savings. 

The combination of cohort masks, policy actions, configuration overrides, and multi-scale outputs is what allows the SPDT to function as a planning tool rather than a single-purpose simulation. A user can compare a heat-pump rollout targeted at social-rent housing against one targeted at the highest consumers, run a retrofit program aimed at households with children alongside a setpoint adjustment for retired households, or test how any of these policies perform under alternative temperature scenarios. The model produces the demand projection; the SPDT exposes the levers that determine what is projected. 

\subsubsection{SPDT}
The SPDT was built to be self-contained and run locally, with the exception of the map-libre map tiles. This simplified the development process and accelerated the prototyping process. The web app is primarily built in python, with html for the and some javascript webpages. It makes use of a number of standard libraries, all of which are open source: 
\begin{itemize}
    \item The web app is built using the flask web application framework. The frontend uses bootstrap5, to ensure responsiveness and allow it to operate on mobile. User inputs are enabled through wtforms, allowing for data validation and consistent rendering. 
    \item The database is built in SQlite, a fast and self-contained database engine. Communication between the database and the web app makes use of SQLAlchemy and associated libraries.  
    \item Data handling within the app is achieved through pandas, geopandas, json, and geojson. The ABM makes use of mesa, mesa-geo, pyarrow and shapely.  
    \item Graphing is provided through plotly, which allows for interactable, web-native figures. GIS functionality is through maplibre-gl, a typescript library which allows for interactive maps. 
\end{itemize}
The application makes use of a number of custom objects, each of which (with the exception of the climate parquet) corresponds to a table in the database. These objects fall into four classes: 
\begin{itemize}
    \item The source data objects contain the source data used by the ABM. These are static and are imported into the database at install.
    \item Scenario objects are user created and contain the information used to initialise the models, as well as other associated metadata such as username and creation date. Each Scenario item exists in an optional one-to-one relationship with a set of record items and a many-to-one relationship with a PolicyChoices object. In both cases, the primary key of the Scenario object is used as the foreign key for the other objects.
    \item Reports objects contain the output of a scenario. They cannot be modified directly by users but can be overwritten by re-running the scenario. 
    \item The policy objects contain information on the policy choices for each scenario. These are user created in the same way as a scenario. 
\end{itemize}

\begin{figure}[htb]
\centering
\includegraphics[width=0.8\textwidth]{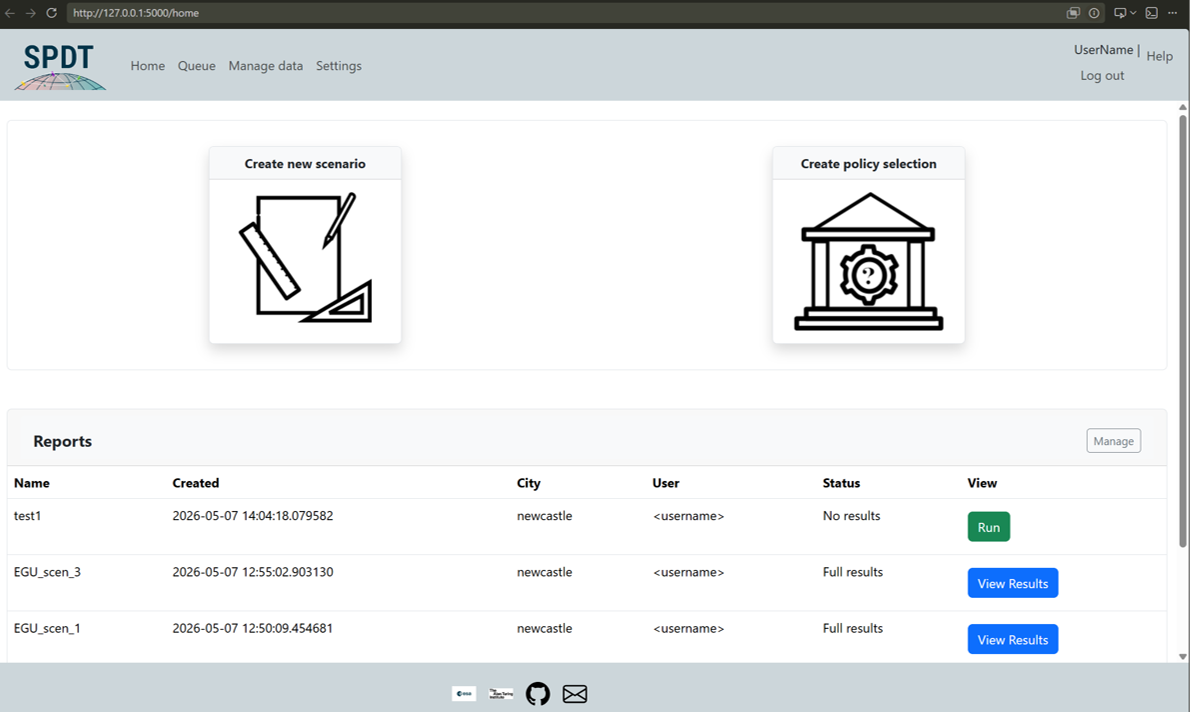}
\caption{\label{fig:homepage} The final version of the home page.}
\end{figure}

The SPDT demonstrator is built to support four primary workflows: creating a scenario, creating a policy selection, running scenarios, and viewing reports. These are accessible via the home page, shown in \Cref{fig:homepage}.

\begin{figure}[htb]
\centering
\includegraphics[width=0.8\textwidth]{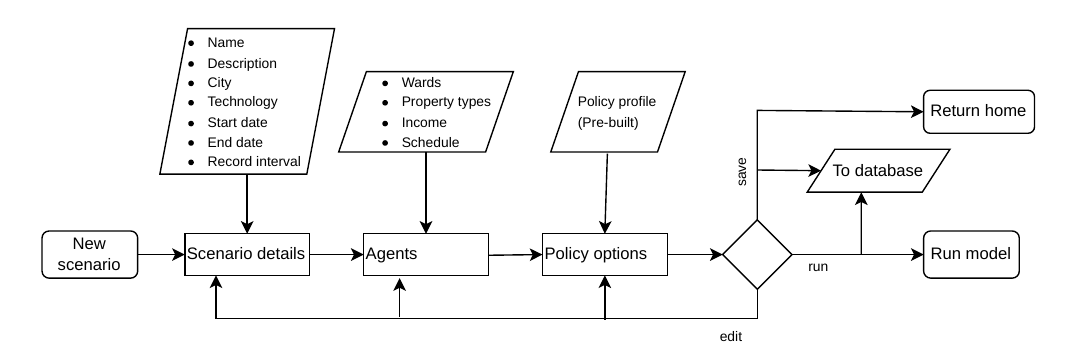}
\caption{\label{fig:WF_newscenario} User workflow for creating a new scenario.}
\end{figure}

Figure \ref{fig:WF_newscenario} shows the primary workflow for creating a scenario. The user is guided through four web pages. At each step, the user input is stored in the session cache. First, they input the object metadata on the scenario details page. This is followed by the agents page, which allows the user to select which agents will be simulated, and the policy options page, which allows the user to select a pre-built set of policy options for the model. Finally, the user gets a summary page that allows them to review their input and choose to edit them, save the model, or run the model. If they select ‘save’ then the app creates a new scenario object using the data in session and then returns the user home. Selecting run will also save the scenario and then direct the user to the scenario status page while the model runs in the background. 

\begin{figure}[htb]
\centering
\includegraphics[width=0.8\textwidth]{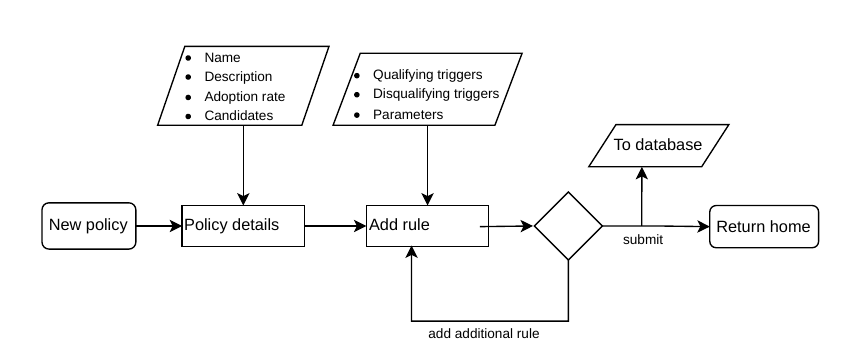}
\caption{\label{fig:WF_newpolicy} User workflow for creating a new scenario.}
\end{figure}

Users can create a new policy selection using the workflow shown in \Cref{fig:WF_newpolicy} on the ‘Create New Policy’ page . This takes place on a single webpage, using two forms. The first form defines the PolicyChoices object, the second defines the Rules objects and can be run multiple times. As with the Scenarios, each rule is stored in the session and shown to the user in the ‘Staged Rules’ box. Once the PolicyChoices form is submitted, a new PolicyChoices item is created in the database along with each of the associated Rules. 

Users can run a scenario immediately after creation or by selecting the option for a previously saved scenario from the home page. In either case, the app follows the workflow shown in Figure 9. First, the appropriate records (as defined by the agents attribute of the scenario) are loaded from the source data tables into a Pandas data frame. The  main heating type parameter of the agents identified by the  PolicyChoices object is then switched to heatpump. The resulting dataframe is used along with the climate data to create a mesa model, which is then run. Each of the report objects are then saved to the database. Once the model has been run and saved, the user is redirected to a ‘success’ page. 

\begin{figure}[htb]
\centering
\includegraphics[width=0.8\textwidth]{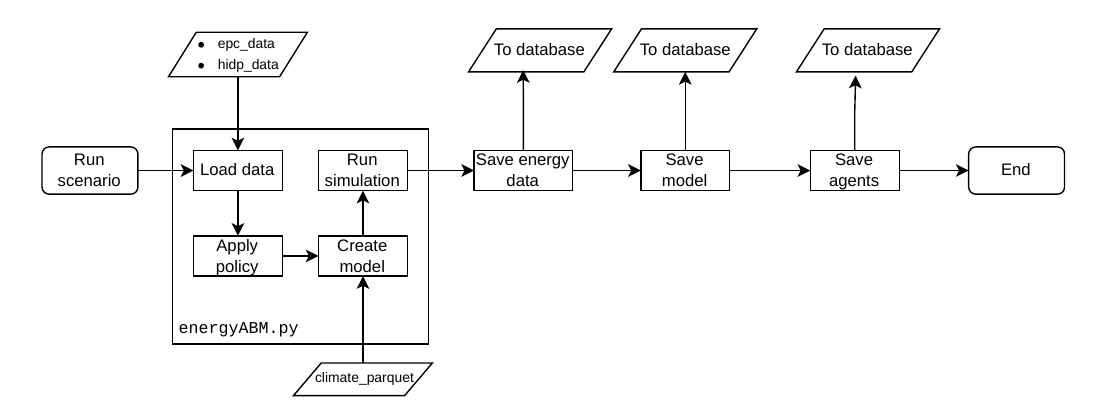}
\caption{\label{fig:WF_run} User workflow for running a scenario.}
\end{figure}

\begin{figure}[htb]
\centering
\includegraphics[width=0.8\textwidth]{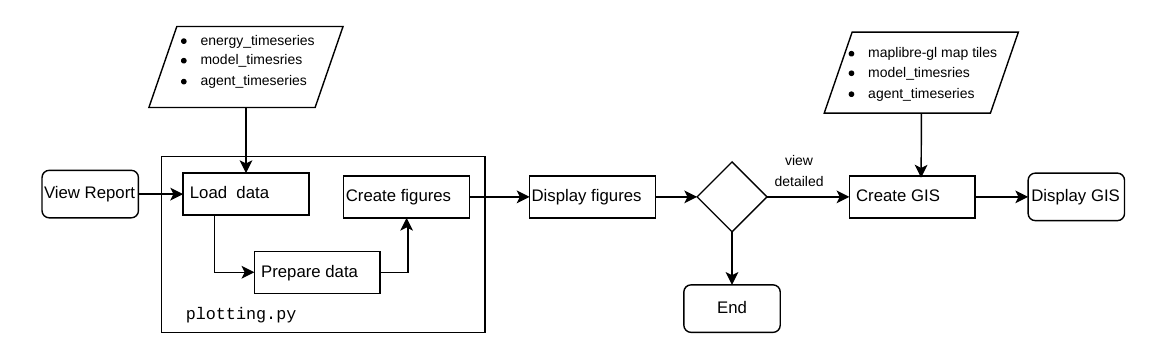}
\caption{\label{fig:WF_view} The final version of an instance of the Summary Report page, including (9) The Summary Information box; (10) The data visualisation tab (with the Heatmap as the active selection); and (11) the hyperlink to the detailed report. Note that agents not included in the scenario are shown in grey.}
\end{figure}

\begin{figure}[htb]
\centering
\includegraphics[width=0.8\textwidth]{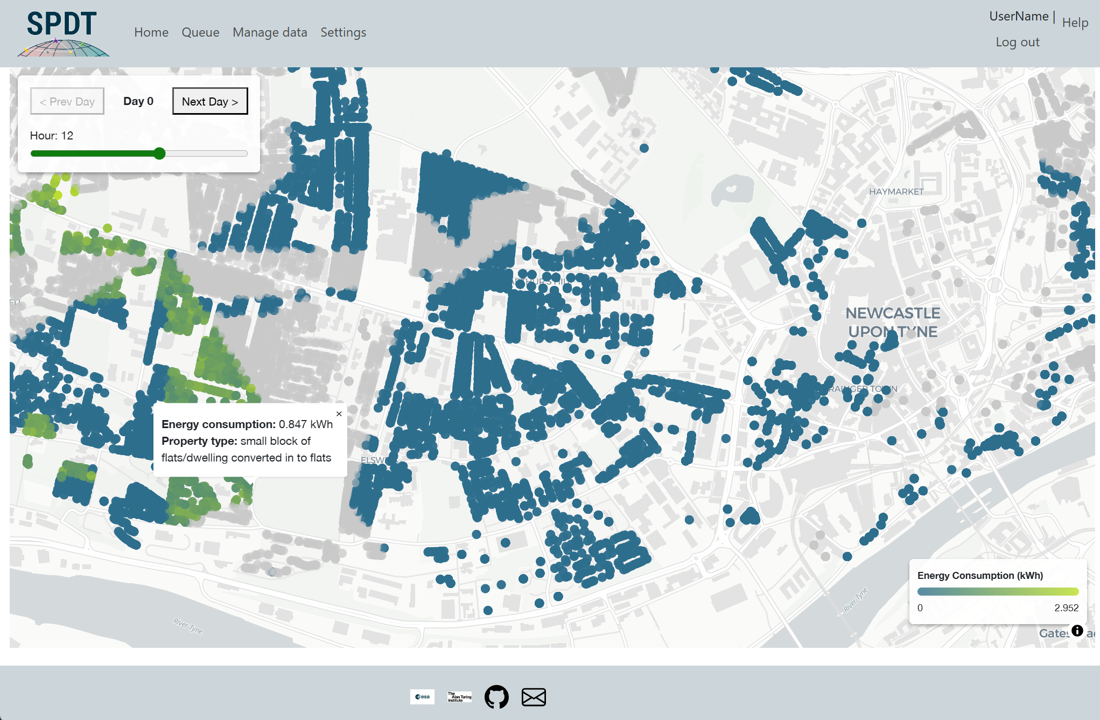}
\caption{\label{fig:report2} An instance of the `detailed report' GIS interface, which enables users to explore the data at each recorded time step.}
\end{figure}

To view a report, the user selects the option from the homepage for a scenario that has all three report objects associated with it. As shown in \Cref{fig:WF_view}, the app then loads the data from these objects, prepares them by resetting the agent IDs and then generates figures using plotly. These are then shown to the user on a reports page, along with some summary information and a ‘detailed timeline’ button. If the user selects this option, the app generates a GIS interface (\Cref{fig:report2}) from the agent and model timeseries and maplibre-gl map tiles and redirects the user to a page displaying it. Users can explore the GIS using the time selection interface and select individual agents to bring up a box containing additional information. 

\section{Discussion \label{sec:discussion}}

In its current form, the case study follows only a single iteration of the design process (see, for example, \cite{Wagg20}). Further integration of the platform into the customer's systems would impose significant additional burden on the NCC, and as such remain out of scope. This limits the end results to technical demonstrator, rather than full digital twin --- live data flows are conspicuously absent, as are more robust data management methods. Nevertheless, the case study serves to demonstrate the early stages (pre-operational) of the design process. Crucially, the case study demonstrates the roles that testing, validation, and verification play in the framework.

In \Cref{sec:challenges}, we described one of the major challenges of policy digital twins being the heterogeneity of each context and ecosystem. This is evident in multiple cases in the case study. Firstly, the context of being a city council in England constrains the stakeholder requirements by shaping both the needs and the possible actions available to NCC. However, this is further compounded by the data availability; the Energy Performance Certificate dataset used is an artifact of the legal framework and data availability in the UK. Its very existence shaped the project --- without this dataset, either substantial additional data collection work would be needed at the outset, or, more plausibly, a different (and potentially inadequate) proxy would have been used.

The specific context in which NCC operates also contributes to this heterogenity. The final review with NCC highlighted a number of additional policy areas and questions which could see the digital twin being applied, including local level energy infrastructure, solar panels on council owned buildings, and charging of council owned electric vehicles. Together, these form a highly specific selection of issues. Although these may overlap with the needs of other councils in similar positions within the country, the specific combination and priority of these issues is likely to be unique. If development were to continue through another iteration of the design process, by incorporating these additional customer requirements, then the digital twin would become increasingly situational, and its applicability to other stakeholders would reduce.

This question of specific utility vs broad applicability is a key determiner in the way that further development would continue and the way in which the platform architecture might evolve. Thus far, it has been assumed that the former should be prioritised --- that is the goal is a digital twin which is maximally useful to a single stakeholder. This would entail expanding beyond the single use case of heat pumps to incorporate these additional design user requirements. A deeper integration into the users' systems (both cyber-physical and organisational) would be expected, allowing the twin to make use of non-public data owned by the user. This will lead to new questions concerning the emergent complexity that arises from the interaction of multiple different models, and require an additional level of robustness in the validation of the operational digital twin, as the combined uncertainty of outputs relying on multiple models comes into play.

However, this is not the only paradigm. A second option is to consider the model as a digital twin component, made available to multiple potential users. This is the paradigm followed by the European Space Agency's DestinationEarth, for example. Rather than working with a single user to create a bespoke platform, instead, a general digital twin architecture is provided, allowing users to pick and choose from multiple components. In this case, the development challenge shifts away from issues regarding multi-model interactions and instead a greater focus is placed on the robustness of both the model and data pipelines for a broad base of users with different contexts. The ability to focus on a single use case supports this, but is accompanied by a loss of control and a looser feedback cycle from the end users. 


The case study highlights the importance of validation and defining the limitations of a policy digital twin throughout its development. Validating across five cities supports a limited claim: the approach transfers within a class of UK local authorities of similar housing stock and data coverage. It cannot extend to a national twin or to a different context without additional efforts. For a policy twin, validation is continuous. The digital twin systems we build need to be able to self correct or, at the minimum, identify and communicate when the data or model are operating outside of their parameters. 

More broadly, for systems such as this to become more widely used there is a need to develop methods to allow regular model re-calibration, similar to data assimilation approaches (as discussed in Section \ref{sec:data_validation}). The calibration used here was successful, but the model will clearly drift as changes to society, infrastructure, economies and the climate occur. Fortunately much of the required data do exist; EPCs are often updated during house refurbishments, and smart-meter data collect information on very short timescales. The main obstacles are firstly that synthetic individuals cannot be directly associated with real individuals so data assimilation processes must work at aggregate levels \citep{lueck_who_2019}, and established assimilation methods do not map cleanly onto discrete, rule-based agent dynamics anyway \citep{wang_data_2015, ward_dynamic_2016, malleson_simulating_2020}. A pragmatic approach for the time being might be to adopt dynamic recalibration --- where model parameters are updated by model states are not --- which has already been shown to be successful in previous ABM studies~\citep{oloo_predicting_2018, oloo_adaptive_2017}.

Whilst data assimilation / dynamic recalibration might correct changes to model parameters, drift will also occur within the population itself. The synthetic population is derived from the 2021 Census, so over the longer time horizons that the SPDT should support the population will change substantially through migration, demographic change, new construction, or even as the results of policy decisions that the tool is designed to support. A fully operational twin would therefore require a dynamic demographic component and periodic re-synthesis of the population as well as the housing stock, beyond the current functionality of DA or calibration methods.



\section{Conclusion \label{sec:conclusion}}

Policy digital twins are a crucial tool for future governance at all levels, from the municipal role played by smart cities, to international projects such as DestiationEarth. However, their implementation faces many challenges. The heterogeneity of each political context makes it difficult to for the designers of a policy digital twin to learn lessons from similar projects in other contexts. Equally, the multi-year timescales involved in policy work require digital twins to include the capability for continuous validation, in order to avoid the virtual representation diverging from the system it represents. In addition, the ways in which people interact with the physical and virtual elements of the systems complicate both the specific implementation of the digital twin, and the conception as to its very definition.

In this paper, we focus on the capability of Multi-level Agent-Based Models (MABMs) to describe this human behaviour as part of a policy digital twin. By representing the human parts of the system as agents with defined characteristics, we can model the interactions between the agents and themselves, and between the agents and other components of the system. By incorporating this at multiple scales, we can more accurately reflect the complexity of real systems, such as cities.

We present a case study of one such implementation of an MABM in a policy digital twin, focused on the city of Newcastle-upon-Tyne. The aim of this digital twin is to enable the policy makers to better target policies relating to energy transition and heat-pump role out. Through integrating an MABM into the digital twin platform, we were able to model the anticipated effects of policy on energy use against the uncertain backdrop of the changing climate.

Through describing this case study we highlight the ways in which situational and policy context influence the design of digital twins. In particular, we describe the impact that the choice of a single stakeholder has on the project. We contrast this with the multi-stakeholder, international approach taken by DestinationEarth, and note the distinction between a specific digital twin implementation, and a general digital twin component. In doing so we highlight the importance of taking such considerations into account from the outset.

\begin{Backmatter}

\paragraph{Acknowledgments}

The authors would like to acknowledge the support of the European Space Agency (ESA) under contract AO/1-12000/24/I-DT “EO Informed Agent Based Models for Digital Twins Applications”. DW is also supported by UKRI grant EP/Y016289/2, and their support is gratefully acknowledged.

We are also grateful to the team at Newcastle City Council for generously providing their time and feedback for the case study.



\paragraph{Competing Interests}
None

\paragraph{Data Availability Statement}

The Multi-level Agent-based Model (MABM) and public datasets discussed in this paper are available on Github: \citet{beltran_2026_21356630}. The Scenario Planning Digital Twin is also available on Github: \citet{mtipuric_2026_21340939}.

\paragraph{Ethical Standards}
The research meets all ethical guidelines, including adherence to the legal requirements of the United Kingdom.

\paragraph{Author Contributions}

\textbf{Matt Tipuric:} Methodology, Software, Data curation, Investigation, Validation, Visualization (equal with AB, NM), Writing -- original draft (lead), Writing -- review \& editing.
\textbf{Alejandro Beltran:} Methodology, Software, Data curation, Formal Analysis, Investigation, Validation, Visualization (equal with MT, NM), Writing -- original draft, Writing -- review \& editing.
\textbf{Nick Malleson: }Conceptualization, Methodology, Visualization (equal with MT, AB), Writing -- review \& editing, Funding acquisition (equal with DA, DW), Supervision (equal with DA).
\textbf{Dani Arribas-Bel:} Conceptualization, Methodology, Writing -- review \& editing, Funding acquisition (equal with NM, DW).
\textbf{David Wagg:} Conceptualization, Methodology, Writing -- review \& editing, Funding acquisition (equal with NM, DA), Supervision (equal with NM).

All authors approved the final submitted draft.


\printbibliography
\end{Backmatter}

\end{document}